\documentclass{bvm} 

\begin{document}

\newcommand{\bvmyear}{2023}

\selectlanguage{english} 

\title{Classification of Prostate Cancer in 3D Magnetic Resonance Imaging Data based on Convolutional Neural Networks}


\titlerunning{Classification of Prostate Cancer based on CNNs}

\author{
	Malte \lname{Rippa} \inst{1, 2},
	Ruben \lname{Schulze} \inst{2},
	Marian \lname{Himstedt} \inst{1},
        Felice \lname{Burn} \inst{3}
}

\authorrunning{Rippa et al.}

\institute{
\inst{1}  Institut für Medizinische Informatik, Universität zu Lübeck\\
\inst{2} FUSE-AI GmbH, Hamburg \\
\inst{3} Cantonal Hospital Aarau, Aarau (CH) \\
}

\email{malte.rippa@gmx.de}

\maketitle

\begin{abstract}
Prostate cancer is a commonly diagnosed cancerous disease among men world-wide. Even with modern technology such as multi-parametric magnetic resonance tomography and guided biopsies, the process for diagnosing prostate cancer remains time consuming and requires highly trained professionals.
In this paper, different convolutional neural networks (CNN) are evaluated on their abilities to reliably classify whether an MRI sequence contains malignant lesions. Implementations of a ResNet, a ConvNet and a ConvNeXt for 3D image data are trained and evaluated.
The models are trained using different data augmentation techniques, learning rates, and optimizers. The data is taken from a private dataset, provided by \emph{Cantonal Hospital Aarau}.
The best result was achieved by a ResNet3D, yielding an average precision score of 0.4583 and AUC ROC score of 0.6214. 
\end{abstract}

\section{Introduction}

With 1,276,106 newly diagnosed cases world-wide in 2018, prostate carcinomas (PCa) are the second most frequently diagnosed cancer disease and account for 3.8\% of all deaths related to cancerous diseases among men \cite{Rawla}. The methods for PCa diagnosis are constantly improving and have reached a new pinnacle with the introduction of multi-parametric magnetic resonance imaging (mpMRI) \cite{Rouviere}. mpMRI describes the usage of multiple imaging sequences such as T2-weighted (T2W), diffusion weighted image (DWI), dynamic contrast enhanced (DCE) and apparent diffusion coefficient (ADC) sequences. Every sequence reveals different characteristics of the abdominal tissue, which allows a broad assessment of the prostate \cite{Rouviere}.
However, the analysis of mpMRI sequences for PCa diagnosis remains a time consuming task and requires further assessment of the severity and clinical significance of the identified lesion(s) e.g. by conducting biopsies.
In the current state of the art in medical image processing, neural networks have been shown to provide reliable predictions of the clinical significance of lesions on different types of images \cite{Castillo}.
It was shown that a convolutional neural network (CNN) can predict the Gleason grade of a histological slice of a prostate biopsy \cite{Karimi}. 
Recently, it was proven that CNNs can reliably detect carcinomas in liver images, when trained with MRI sequences and histopathological ground truth \cite{Oestmann}.

The scope of this paper is to compare the performance of different CNN architectures, that predict whether a prostate contains malignant lesions, based on whole mpMRI sequences and information gathered from histopathological tissue assessment as image level ground truth label during training. In contrast to state of the art methods \cite{Li} the use of voxel-wise annotations is omitted as it is to be determined whether a whole image classification can be realized reliably. A whole image classification model would profit from a larger amount of data as image level labels are easier to obtain, perhaps resulting in a more robust and stable classifier. In addition a well performing classifier could be used as a filter without the need to infer computationally expensive segmentation models.

\section{Material and Methods} 
\subsection{Image Data}
The data for the experiments was taken from a private dataset exclusively provided by \emph{Kantonspital Aarau}, Switzerland.
The dataset is structured in cases, studies and sequences, where a case contains one or more studies and a study contains one or more mpMRI sequences. Studies that contain a T2W image of insufficient quality because of artifacts that hide the prostate in the image e.g. due to endorectal coils, hip implants or anatomical phenomena (bladder protruding into the prostate or similar) were excluded from the dataset. After refinement a dataset consisting of 365 studies and 1095 image sequences has been obtained. Each image has a size of $149\times149\times32$ with a voxel spacing of 0.75, 0.75, 3 in x, y, z direction, respectively.

\subsection{Labels}
For each study, a histopathological report is available, containing the Gleason score per prostate sextant. The Gleason grading system is considered one of the most powerful grading systems in prostate cancer analysis. It provides information about the condition of the tissue \cite{Gordetsky}. The information from the reports are refined to image level binary labels, by condensing the Gleason scores into one binary score. Per definition of the scoring system \cite{Gordetsky}, a Gleason score of  $\geq 3+4=7$ is considered as malignant/clinically significant. Therefore, if one sextant contains a malignant lesion, the entire image is labeled with 1, else 0 for indicating benign/clinically insignificant lesions or no lesions at all.
Inconsistencies, errors and incompleteness in human annotation make the ground truth unreliable or incomplete for some cases. The entire study including the image data was excluded from the dataset, if the ground truth is found to be of insufficient quality due to aforementioned reasons. This resulted in a total of 246 benign labels and 119 malignant labels.

\subsection{Models}
Generally, CNNs are considered appropriate for fast image processing and are widely used in computer vision tasks. It is suggested by literature \cite{Castillo, Oestmann} to use CNNs for the classification of prostate cancer in mpMRI data by delivering strong results when applying CNNs for similar problems.
For this paper the ConvNet3D, was chosen due to its simplicity combined with efficiency. A ResNet3D was chosen as the use of residual connections could be beneficial. Recently, the ConvNeXt was published. It is described as modernized ResNet incorporating the principles of transformers and is reported to deliver strong performance for object detection.
All these models rely on a CNN backbone and use fully connected layers for the final classification.
The ConvNet3D uses two convolutional blocks to increase the channel/feature dimension while decreasing the spatial resolution. It has 168,705 trainable parameters.
The ResNet3D consists of eight convolutional blocks containing convolution operations, batch normalization and ReLU. The residual connections are realized by using “bottleneck blocks”, which are convolutional blocks that add the output after one convolution block to the input before the next convolution block. Six bottleneck blocks are used. In total 4,527,906 parameters have to be learned.
The ConvNeXt3D uses features such as grouped convolutions, inverted bottlenecks with modified normalizations and activation functions. Using a combination of three convolution blocks, three inverted bottleneck blocks containing three convolutional layers each and a fully connected layer results in 31,321,561 trainable parameters.

All models output a single class activated using sigmoid, indicating the confidence of containing a malignant lesion.

As baseline, the institutional inhouse lesion segmentation model that is currently deployed in a product by \emph{FUSE-AI} 
was run on the same dataset. The model, an anisotropic U-Net, is supposed to create a segmentation  for benign and malignant lesions. It therefore must be able to classify the clinical significance of detected lesions. For comparability to the other models mentioned earlier, the model output is aggregated using global max pooling to generate a classification score.

\subsection{Training process and experiment setup}
Selecting only data with sufficient image and ground truth quality leaves 365 cases with a total of 1095 sequences for training. Three MRI sequences (T2W, ADC and DWI) are stacked in the channel dimension to compose a three channel 3D image as input for the model. It was proven that incorporating DWI and ADC sequences leads to increased performance when aiming to recognize anomalies in the prostate \cite{Li}. 

The data is then split into a training, test, and validation partition by 70, 15, and 15 percent, respectively. To improve generalization, data augmentation in form of mirror transforms, changes in contrast and resolution, addition of noise, and spatial transforms is applied to the images.

\begin{figure}[!h]
\centering
\includegraphics[width=0.4\linewidth]{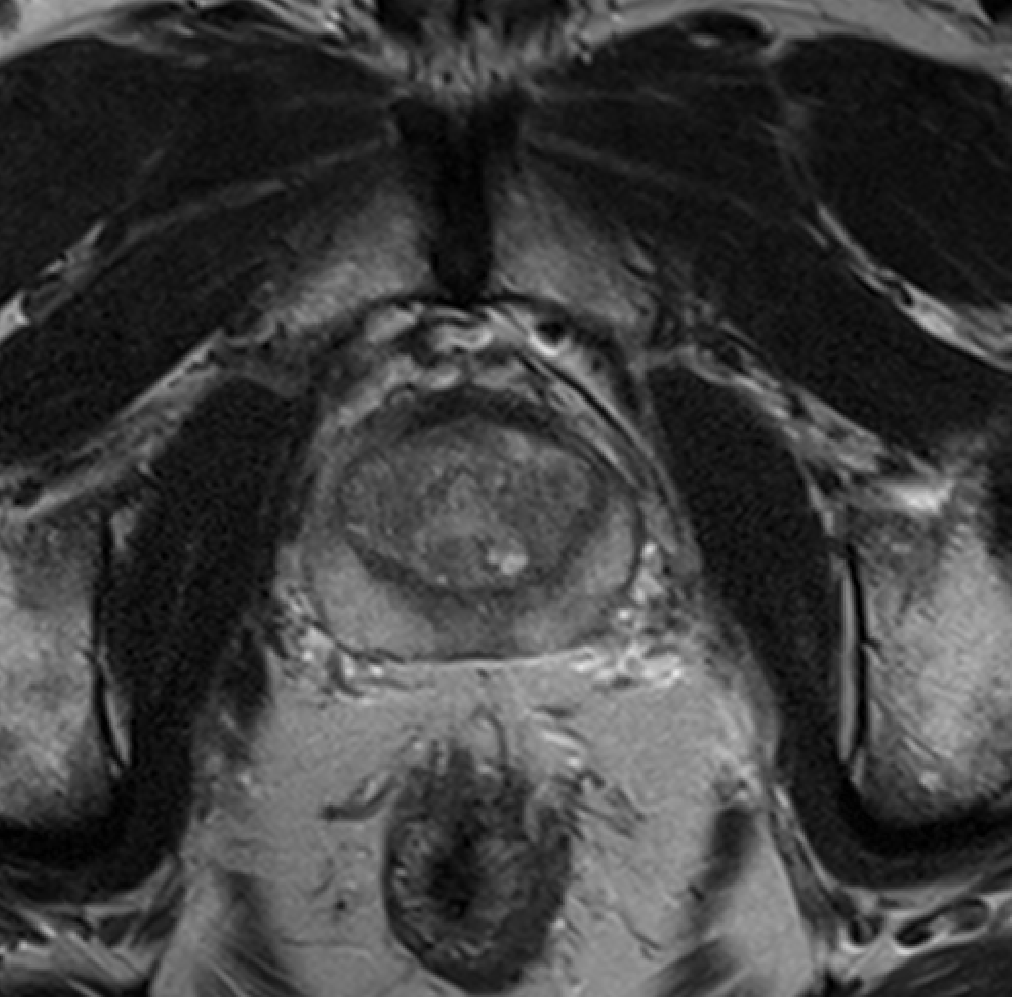}
\includegraphics[width=0.4\linewidth]{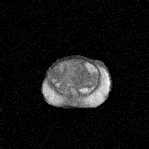}
\caption{Left: The raw input image (here T2W), Right: The cropped prostate as image input for the model}

\label{fig:prostate_cropped}
\end{figure}

To enhance the focus on the important regions of the image, an existing model is utilized to generate a segmentation of the prostate (see Fig. \ref{fig:prostate_cropped}, that is then passed to the classification network for further processing.
The selected models are applied to the whole image as well to measure the impact of the pre-segmentation. However, only AUC ROC is calculated for the whole image models.
The models were trained multiple times with different hyperparameters and loss calculations, to encounter a training setup that delivers reasonable results. Once the results showed that the performance of the networks was stable, the setup was considered successful and fine tuning of the hyperparameters could commence. To counter the imbalances in class distribution, the binary cross entropy (BCE) loss was weighted with inverse class frequencies.
Additionally, the networks were trained on different partitions of the entire dataset and smaller subsamples of the entire dataset (40 to 50 samples), to determine if the composition of data impacts the outcome of the model training as no cross validation was conducted.
The performance of the different network implementations was measured by calculating the area under the curve of the receiver operating characteristic (AUC ROC), the average precision (AP) and the analysis of confusion matrices.

As reference the performance of a UNet-based lesion segmentation pipeline is used denoted as "Baseline model" in Tab. \ref{tab:results}. The results of the lesion segmentations are aggregated to compose an image-level label.

\section{Results} 

It can be seen directly, that the introduction of pre-segmentations around the prostate gland caused a positive effect on the AUC ROC score. In Tab. \ref{tab:results} it is documented that the models on the whole image perform with an average AUC ROC of 0.5075 whereas the models that focus only the gland yield an average AUC ROC of 0.5970.
The ResNet3D was trained for 300 epochs with an initial learning rate of $8\mathrm{e}{-6}$ and exponential and weight decay by $1\mathrm{e}{-4}$ every 100 steps, with a batch size of 2, a dropout rate of 70\% and an AdamW optimizer. It was observed that the network started to overfit after approximately 19 epochs. Decreasing the initial learning rate prevented overfitting but led to worse results for the validation and test partition.

The ConvNet3D was trained in the same setup as the ResNet3D. After fine tuning the learning rate to $1\mathrm{e}{-6}$, the AUC ROC score and AP did increase slightly to 0.6214 and 0.4583 respectively in the best epoch for the test partition.
The ConvNeXt3D overfitted in the first epochs on an initial learning rate of $8\mathrm{e}{-6}$, as well, however further fine tuning did not improve the results.

It is to mention, that the training with smaller learning rates did not lead to any learning at all. The loss of the classification of the training data is oscillating throughout the entire training of 300 epochs and thus the scores for AUC ROC and AP are not improving over time. The best performance of this network was achieved in the earlier epochs of the training and the score decreases over time, while the loss remains constant.

As can be seen in Tab. \ref{tab:results} the best result was achieved by the ConvNet with the prostate segmentation as input in terms af AUC ROC, yielding 0.5732. At the same time the model yields the worst AP. In terms of AP none of the models was able to outperform the baseline model, yielding an AP of 0.4801 at best.

\begin{SCtable}[10][h]
\begin{tabular}{lcc}
\hline
Model & AUC ROC & AP \\
\hline
Baseline model & \textbf{0.6343} & \textbf{0.6967}\\
ResNet3D (Whole image) & 0.4832 & / \\
ConvNet3D (Whole image) & 0.4999 & / \\
ConvNeXt3D (Whole image) & 0.5395 & / \\
ResNet3D 		& 0.6214 & 0.4583\\
ConvNet3D 			& 0.5964 & 0.3743   \\
ConvNeXt3D  & 0.5732 	& 0.4801 \\
\hline
\end{tabular}
\caption{Performance of the different networks (all entries for the respective best test epoch)}
\label{tab:results}
\end{SCtable}


For none of the models a significant difference in applying weighted BCE or regular BCE loss could be observed.

Training the models on small subsets of the dataset, delivers strong models with AUC ROC and AP at 1 for training and over 0.8 for validation and test.
The results are not reliable, however, as discussed later.

\begin{figure}[!h]
\centering
\includegraphics[width=5.5cm]{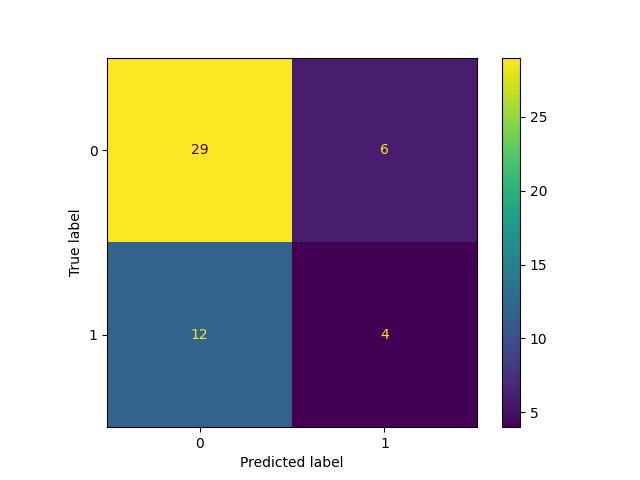}
\includegraphics[width=5.5cm]{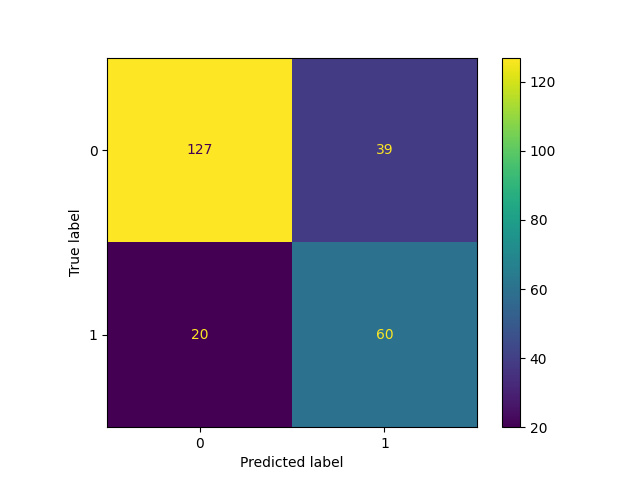}

\caption{Confusion matrices for the binary classification of prostate cancer with the ResNet3D after 300 epochs, Left: test partition, Right: training partition}
\label{fig:cm}
\end{figure}
\section{Discussion}
Several different convolutional neural networks were tested in different training configurations to obtain the best performance for each network. The performance was measured by calculating AUC ROC and AP for training, test and validation. The test scores were used for comparison of the models. The scores were comparably low, with an AUC ROC of 0.6214  and AP of 0.4583 at best. In terms of AUC ROC the ResNet3D performed similar to the inhouse solution on this dataset, but the low AP shows that predictions are not precise. Fig. \ref{fig:cm} shows the confusion matrices for training and test after 300 epochs of training. Analyzing the confusion matrices, the majority of correctly classified samples are benign (clinically insignificant) cases. This adheres directly to the distribution of the class labels. Also the high rate of false positives and false negatives leads to decreasing scores considering AUC ROC and AP.

It can be observed that the different models do not generalize well on small learning rates in the here presented setup. Usually, the best epochs can be found among the first training epochs.
Increasing the learning rate leads to overfitting, which delivers a model that learned the distribution of the training data instead of learning to find a generalized representations of the input. Furthermore, the good performance of the models on smaller subsamples of the dataset, support the observation of the networks learning distributions instead of representation. Especially on the smaller sets, it is more likely to randomly select a set for which the distribution of training data matches the distribution of validation and test data, which results in perfect but unreliable results. The subsamples are very small considering that a median dataset size of 127 patients/studies is reported when comparing other approaches on PCa classification on MRI data \cite{Castillo}.

The results show that the chosen networks are not able to provide reliable predictions of the clinical significance of lesions in the prostate by processing the MRI sequences without additional information or improvements in the training process. 
In literature an AUC ROC of 0.8328 $\pm$ 0.0878 is reported for lesion detection and characterization \cite{Li}, which sets the lower boundary to be reached by the model proposed in this work. It is to mention that the well performing models in related work use lesion locations \cite{Aldoj} or lesion segmentations as ground truth for training a UNet \cite{Arif}. Making use of pretrained larger ConvNets, e.g. Inception v3 \cite{Armato} in combination with zonal information and lesion locations delivers strong results on similar input as proposed in this paper.

In future work, more than one data source should be considered, which would help concluding if the data quality is insufficient or if the hyperparameters are chosen poorly. Additionally, it would be useful to assess other machine learning algorithms such as logistic regression, SVMs or visual transformers, instead of focusing exclusively on CNN architectures. 
Moreover, the implementation of cross validation could help to measure the performance more accurately.
As suggested by the state of the art and the results achieved in this paper,  incorporation of localization information in form of anatomy segmentations should be targeted. It could be helpful to use the entirety of information from the histopathological report instead of the aggregated label, thus incorporating sextant information. Using pretrained weights and/or self-supervised training was proven as effective \cite{Armato,Zhou2020} and should be tried as well.

\begin{acknowledgement}
The work has been carried out at FUSE-AI GmbH in Hamburg and supervised by the Institute of Medical Informatics, Universität zu Lübeck.
Thanks to Alexander Cornelius, Sebastian Schindera, Rainer Grobholz, Stephan Wyler and Maciej Kwiatkowski from \emph{Kantonspital Aarau} for providing the image data and histopathological reports.
Thanks to Quang Thong Nguyen for the extensive help on the preparation of the dataset and the implementation of ResNet3D, ConvNet3D and ConvNeXt3D.

\end{acknowledgement}

\printbibliography

\end{document}